# Quantum-Enhanced Recurrent Neural Networks via Variational Quantum Gating for Battery State of Health Prediction

Yin Xu[1], Qinglin Liu[2], Li Gao[2] and Hua Xu[1,3]*

[1]College of Artificial Intelligence, Tianjin University of Science and Technology, Tianjin 300457, China

[2]Chery Automobile Co., LTD., Wuhu City, Anhui Province 241000, China

[3]Yiwei Quantum Technology Co., Ltd, Hefei 230088, China

Corresponding Author:

H. Xu

Email: hua.xu@ywquantum.com

**Abstract**

Accurate state-of-health (SOH) estimation for lithium-ion batteries remains a challenging problem due to complex electrochemical degradation mechanisms and long-range temporal dependencies. In this work, we propose a quantum-enhanced recurrent framework, termed QLSTM, in which variational quantum circuits are directly embedded into the gating mechanisms of long short-term memory networks. By replacing classical affine transformations with parameterized unitary operations, the proposed model introduces structured nonlinear transformations into the recurrent state-transition process. Extensive experiments on multiple benchmark battery datasets demonstrate that QLSTM consistently outperforms classical sequence models in both predictive accuracy and robustness, achieving significant reductions in mean absolute error (MAE), with improvements on the order of 20% compared with classical LSTM baselines. Ablation studies further confirm that these improvements arise primarily from quantum-enhanced gating rather than input-level transformations. Additional analyses on qubit scaling and noise robustness reveal that model performance is governed by a balance between expressive capacity and trainability. These results provide empirical evidence that embedding quantum computational primitives within recurrent architectures offers a structurally grounded approach to improving sequence modeling capability. The proposed framework establishes a new design paradigm for integrating quantum operators into temporal learning models, with potential applications in complex dynamical system prediction tasks.

1. Introduction

Lithium-ion batteries play a central role in electric vehicles (EVs) and renewable energy storage systems. Their degradation is governed by complex electrochemical mechanisms, including solid electrolyte interphase (SEI) growth, loss of active material (LAM), and loss of lithium inventory (LLI) [1]. These processes exhibit strong nonlinear coupling and long-term temporal dependencies, making accurate state-of-health (SOH) estimation a challenging sequence modeling problem. Reliable SOH prediction is therefore essential not only for preventing thermal runaway and extending battery lifetime, but also for enabling proactive battery management systems (BMS).

Existing SOH estimation methods can be broadly categorized into model-driven and data-driven approaches. While model-driven methods provide strong physical interpretability, data-driven methods—particularly recurrent neural networks such as long short-term memory (LSTM) networks — have shown strong capability in capturing temporal degradation patterns from battery aging trajectories [2]. Nevertheless, most classical recurrent architectures still rely on gating mechanisms constructed from affine transformations followed by fixed nonlinear activation functions. Such designs may limit the expressive capacity of recurrent gating mechanisms in modeling the high-order, highly coupled nonlinear dynamics underlying battery degradation, particularly in heterogeneous electrochemical systems [3], [4], [5].

Quantum machine learning (QML) provides a fundamentally different computational paradigm for addressing these limitations. By encoding classical inputs into quantum states and applying parameterized unitary transformations, variational quantum circuits (VQCs) enable expressive nonlinear mappings in high-dimensional Hilbert spaces, facilitating structured feature interactions beyond classical affine transformations [6], [7], [8]. With ongoing advances in noisy intermediate-scale quantum (NISQ) hardware, hybrid quantum–classical models have become increasingly feasible for practical applications [9], [10], [11]. Recent studies have begun to explore QML-based approaches for battery SOH estimation, suggesting that quantum-enhanced modeling may offer a promising avenue for overcoming representational bottlenecks in classical deep learning. However, most existing approaches remain focused on feature-level quantum transformations, with limited integration into the internal dynamics of sequence models.

Despite this progress, two key challenges remain. First, existing QML-based approaches predominantly focus on static or spatial feature transformations, while the integration of quantum transformations into temporal gating mechanisms for sequence learning remains largely unexplored. Battery degradation is inherently a sequential process that requires recursive state updates and long-term memory retention. This fundamentally distinguishes it from the static classification tasks in which QML has been more extensively applied. Second, most prior studies have been evaluated on single-chemistry datasets, limiting a rigorous assessment of model robustness and generalization. Systematic validation of heterogeneous electrochemical systems一 including different chemistries, cycling protocols, and degradation rates一remains insufficient, making it difficult to disentangle genuine structural advantages from dataset-specific effects.

Motivated by these limitations, we propose a quantum-enhanced recurrent framework, termed Quantum Long Short-Term Memory (QLSTM), for battery SOH prediction. The proposed architecture embeds VQCs into LSTM gating mechanisms, replacing classical affine transformations with parameterized unitary operations. By integrating quantum transformations directly into the recurrent gating, QLSTM enables structured nonlinear feature interactions while preserving the memory dynamics essential for sequential modeling. To systematically evaluate the proposed approach, experiments are conducted on three benchmark datasets spanning distinct lithium-ion battery chemistries, including the MIT dataset based on lithium iron phosphate (LFP) cells, the CALCE dataset based on lithium cobalt oxide (LCO) cells, and a dataset of commercial nickel cobalt aluminum oxide (NCA) cells, with strict cell-level partitioning to ensure unbiased generalization assessment. Furthermore, model robustness is examined under hardware-inspired conditions by incorporating a bit-flip noise model to simulate quantum decoherence across varying noise levels [12], [13].

The main contributions of this work are summarized as follows.

(1) Quantum-Enhanced Recurrent Modeling Framework. We propose a quantum-enhanced recurrent architecture, instantiated as QLSTM, in which variational quantum circuits are embedded into LSTM gating mechanisms to replace classical affine transformations. This design enables structured nonlinear feature interactions within recurrent updates and provides a principled framework for integrating quantum transformations into sequence modeling.

(2) Cross-Chemistry Generalization Capability. We demonstrate that the proposed QLSTM architecture achieves consistently improved predictive performance and robust generalization across heterogeneous electrochemical systems. Through cross-cell and cross-chemistry validation on multiple benchmark datasets, the results reveal the structural advantage of quantum-enhanced gating mechanisms in capturing complex degradation dynamics beyond dataset-specific conditions.

(3) Robustness Characterization under NISQ Constraints. We provide a systematic characterization of quantum-enhanced recurrent models under realistic hardware-inspired conditions by incorporating a bit-flip noise model. The results provide empirical insights into the stability of quantum-enhanced gating mechanisms and their practical feasibility in the NISQ regime.

The remainder of this paper is organized as follows. Section 2 reviews the related work. Section 3 presents the proposed methodology and the feature engineering pipeline. Section 4 reports the experimental results and provides a comprehensive performance analysis. Section 5 concludes the paper and discusses directions for future research.

**2. Related Work**

2.1 Battery SOH Estimation Methods

SOH estimation of lithium-ion batteries has been widely investigated using both model-driven and data-driven approaches. Model-driven methods, such as pseudo-two-dimensional (P2D) electrochemical models [14] and equivalent circuit models (ECMs) [15], provide strong physical interpretability by explicitly modeling internal electrochemical processes. However, these approaches typically require complex parameter identification and calibration, and their performance may degrade under varying operating conditions and battery chemistries. Comprehensive review studies have further highlighted the practical limitations of model-driven approaches [18].

Data-driven methods have gained increasing attention due to their ability to learn degradation patterns directly from data. Early studies employed machine learning techniques such as support vector regression and ensemble learning, while more recent works have adopted deep learning architectures, including convolutional neural networks (CNNs) and recurrent neural networks (RNNs), for SOH estimation [19], [20], [21], in particular, LSTM networks have demonstrated strong capability in modeling temporal dependencies in battery aging trajectories [22]. For example, Park et al. addressed capacity regeneration phenomena using multi-channel charging profiles [23], while Fan et al. modeled non-stationary degradation trends through CEEMDAN-based signal decomposition within an SVR-LSTM hybrid framework [24]. In addition, hybrid approaches combining signal decomposition techniques with deep learning models have been proposed to improve predictive accuracy under non-stationary conditions. Despite these advances, existing data-driven methods still rely on predefined parametric transformations, which may limit their ability to capture the complex, highly coupled nonlinear dynamics underlying battery degradation, especially under heterogeneous operating conditions.

2.2 Deep Sequence Modeling for Battery Degradation

Given the inherently sequential nature of battery degradation, considerable effort has been devoted to improving sequence modeling techniques for SOH prediction.

Variants of recurrent architectures, such as gated recurrent units (GRUs) and enhanced LSTM-based models, have been proposed to better capture long-term dependencies and mitigate issues such as vanishing gradients. More recent studies have explored attention mechanisms, multimodal learning, and hybrid architectures to improve predictive performance under complex and non-stationary operating conditions.

Although these approaches have achieved notable improvements in predictive accuracy, most sequence models still rely on predefined gating structures based on affine transformations and fixed nonlinear activations to perform recursive state updates. Such designs may limit their ability to fully capture complex temporal dependencies and long-term evolution patterns in battery degradation, particularly under heterogeneous operating conditions and varying degradation rates. As a result, enhancing the expressiveness of recurrent gating mechanisms remains a key challenge for accurate SOH estimation.

2.3 Quantum Machine Learning and Hybrid Models

Recent advances in quantum machine learning (QML) have introduced new possibilities for enhancing model expressiveness through quantum-enhanced representations. Variational quantum circuits and related quantum neural network models have been widely studied for realizing nonlinear mappings in high-dimensional Hilbert spaces. With the rapid development of NISQ hardware, hybrid quantum–classical algorithms have become increasingly viable for near-term applications.

In the context of battery SOH estimation, recent studies have begun to explore

QML-based approaches. Variational quantum algorithms have been applied to improve ensemble learning strategies [25], while quantum convolutional neural networks (QCNNs) have demonstrated parameter-efficient feature extraction through quantum encoding [26]. Hybrid quantum–classical frameworks combining quantum models with recurrent architectures have also been proposed to enhance nonlinear modeling capability under varying operating conditions. Preliminary studies have further explored quantum extensions of recurrent architectures [27]; however, their application to battery degradation modeling remains limited, and systematic evaluation across heterogeneous electrochemical systems is still lacking.

Despite these advances, existing QML-based approaches are primarily limited to feature-level quantum transformations, where quantum models are used as external feature extractors or auxiliary modules. The integration of quantum transformations into the internal dynamics of sequence models—particularly within recurrent state-transition and gating mechanisms—remains largely unexplored. This gap motivates the development of quantum-enhanced recurrent architectures that directly incorporate quantum transformations into temporal modeling processes.

## 3. Methodology

### 3.1 Problem Formulation

This study formulates SOH estimation as a nonlinear sequence regression problem over battery degradation trajectories. Let $\mathbf{x}_t \in \mathbb{R}^d$ denote the health indicator (HI) vector extracted at cycle t, where d is the feature dimension after preprocessing and feature selection. Given a historical sequence of length k, the objective is to learn a mapping:

$$f_\theta:(\mathbf{x}_{t-k+1},\ldots,\mathbf{x}_t) \mapsto \hat{y}_t \tag{1}$$

where $\hat{y}_t \in \mathbb{R}$ is the predicted SOH at cycle t, and $\theta$ denotes the model parameters. The learning objective is defined by minimizing the mean squared error (MSE):

$$L(\theta)=\frac{1}{N}\sum_{i=1}^{N}(y_i-\hat{y}_i)^2 \tag{2}$$

where $y_i$ is the ground-truth SOH and N is the number of training samples. Since SOH evolution exhibits both temporal dependency and nonlinear dynamics, an effective estimator must capture long-range sequential patterns as well as complex feature interactions.

Recurrent neural networks, particularly LSTM networks, are well suited to modeling such temporal dependencies through gated memory mechanisms. However, in classical LSTM, state transitions rely on affine transformations followed by element-wise nonlinearities, which may limit the modeling of structured feature interactions in strongly coupled degradation dynamics.

To address this limitation, we adopt a hybrid quantum–classical recurrent formulation in which the affine transformations in LSTM gating mechanisms are replaced by variational quantum circuits (VQCs). From a functional perspective, VQCs can be viewed as parameterized nonlinear transformations that enable structured feature interactions beyond classical linear projections.

The resulting framework preserves the recursive memory structure of LSTM while introducing enhanced flexibility in feature transformation. The overall pipeline, illustrated in Fig. 1, integrates feature engineering, data preprocessing, and quantum-enhanced recurrent modeling into a unified framework for SOH estimation.

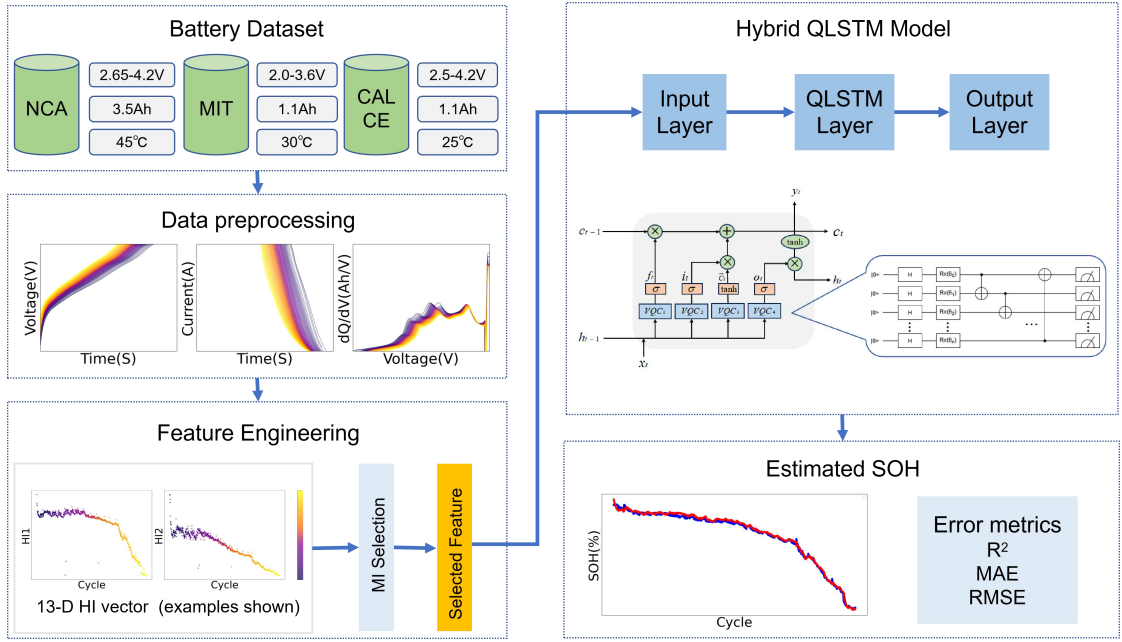

Fig. 1. Pipeline of the hybrid QLSTM framework for battery SOH prediction, covering multi-chemistry data collection, MI-based feature selection, VQC-embedded gating architecture, and multi-metric evaluation.

### 3.2 Feature Engineering

Accurate SOH prediction depends not only on model architecture but also on the quality, physical interpretability, and statistical relevance of the input features. In this study, a physics-informed feature engineering strategy is designed to incorporate electrochemical degradation knowledge into the model input space. This approach is particularly important for hybrid quantum–classical models, in which input dimensionality and feature redundancy can affect the efficiency and stability of VQCs.

The feature engineering pipeline consists of three main steps: (i) computation of the SOH ground truth, (ii) extraction of HIs, and (iii) mutual information-based feature selection.

### 3.2.1 SOH Ground-Truth Computation

SOH is a widely used diagnostic metric for quantifying battery degradation and reflects the cumulative effects of aging mechanisms [16]. It is defined as:

$$SOH_t = \frac{Q_{curr,t}}{Q_{nom}} \times 100\% \tag{3}$$

where $Q_{curr,t}$ is the maximum dischargeable capacity at cycle t, obtained through ampere-hour (Ah) integration of the discharge current over a complete cycle, and $Q_{nom}$ denotes the nominal rated capacity. To ensure numerical stability and consistency across datasets, SOH trajectories are computed exclusively from full discharge cycles, thereby avoiding potential bias from partial cycling or abnormal measurements. In practical applications, a battery is typically considered to have reached end-of-life (EOL) when its SOH declines to 80% [17].

### 3.2.2 Health Indicator Extraction

To characterize battery degradation behavior, thirteen HIs are extracted from charge–discharge profiles and grouped into four categories according to their physical interpretation.

**Time and Operating Condition Features (HI1–HI4):** These features characterize the temporal dynamics of the charging process, including the durations of the constant current (CC) and constant voltage (CV) phases. As degradation progresses, increased internal resistance and polarization alter these time-related patterns, reflecting changes in electrochemical kinetics.

**Energy Evolution Features (HI5–HI6):** These features are computed via

numerical integration over charging segments, providing an estimate of energy utilization and dissipation. From a thermodynamic perspective, they capture the gradual decline in energy conversion efficiency associated with battery aging [18].

**Curve Morphology and Statistical Features (HI7–HI10):** Voltage–time curves exhibit structural distortions during degradation. Statistical descriptors such as slope, skewness, and information entropy are extracted to quantify variations in curve shape, irregularity, and asymmetry, providing an indirect characterization of nonlinear degradation.

**Incremental Capacity Analysis Features (HI11–HI13):** Incremental capacity analysis (ICA) curves are derived from CC charging sequences using Savitzky–Golay smoothing and numerical differentiation [28]. The peak voltage, magnitude, and area of the dominant ICA peaks are extracted, reflecting capacity fading associated with electrochemical processes such as LAM and LLI [29].

The extracted HIs provide a structured representation of battery aging from multiple perspectives, including kinetic, energy, and phase-transition characteristics. The detailed definitions and physical interpretations of all HIs are summarized in Table 1.

Table 1. Summary of extracted HIs and their physical interpretations.

| Symbol | Feature Description | Physical Interpretation |
|---|---|---|
| HI1 | CC charging time [s] | Reflection of internal resistance and polarization shifts |
| HI2 | CV charging time [s] | Decline in charge acceptance capability |
| HI3 | Current at 200s of the CV phase [A] | Current decay characteristics reflecting kinetic degradation |
| HI4 | Voltage at 500s of the charging process [V] | Dynamic response characteristics under aging-induced overpotential |

| HI5 | Energy during the CC-CV charging segment [Wh] | Characterization of energy utilization efficiency |
| HI6 | Energy ratio of CC to CV segments | Evolution of power distribution during the charging process |
| HI7 | Terminal voltage at charging onset [V] | Correlation with initial polarization and OCV recovery. |
| HI8 | Slope of the CC voltage-time curve | Variations in polarization impedance during the CC phase |
| HI9 | Information entropy of the CC voltage curve | Degree of signal irregularity and complexity |
| HI10 | Skewness of the CC voltage curve | Asymmetry of the curve morphology due to nonlinear aging |
| HI11 | Peak voltage of the dQ/dV curve [V] | Voltage shift associated with electrode phase transitions |
| HI12 | Peak area of the dQ/dV curve | Capacity loss resulting from LAM and LLI |
| HI13 | Peak magnitude of the dQ/dV curve | Intensity variation of phase transitions during lithiation |

### 3.2.3 Mutual Information-Based Feature Selection

Given the sensitivity of VQCs to input dimensionality and qubit scaling, feature selection is performed to reduce redundancy while preserving informative degradation signals. In this work, a dual-metric evaluation strategy is adopted, combining mutual information (MI) and Spearman's rank correlation coefficient ($\rho_s$) [30], [31]. MI is used as the primary selection criterion, as it measures the statistical dependency between each feature and the SOH target, capturing both linear and nonlinear relationships. Spearman's $\rho_s$ serves as a complementary metric to characterize monotonic trends between variables.

As illustrated in Fig. 2, certain features exhibit relatively low rank-correlation coefficients but comparatively high MI values across datasets, such as HI5 in the NCA dataset and HI12 in the CALCE dataset. This observation indicates that these features preserve nonlinear relationships with SOH that may not be captured by correlation-based measures alone, thereby supporting the use of MI as the primary selection criterion [32].

To prevent information leakage, feature ranking and selection are performed exclusively on the training partition and conducted independently for each dataset to account for chemistry-specific degradation behavior. Based on the MI scores, the top 10 features are retained to construct the input vector, while features with minimal informational contribution are discarded. This selection strategy reduces input dimensionality and aligns the feature space with the representational capacity of VQCs, facilitating efficient nonlinear feature transformation under constrained quantum resources.

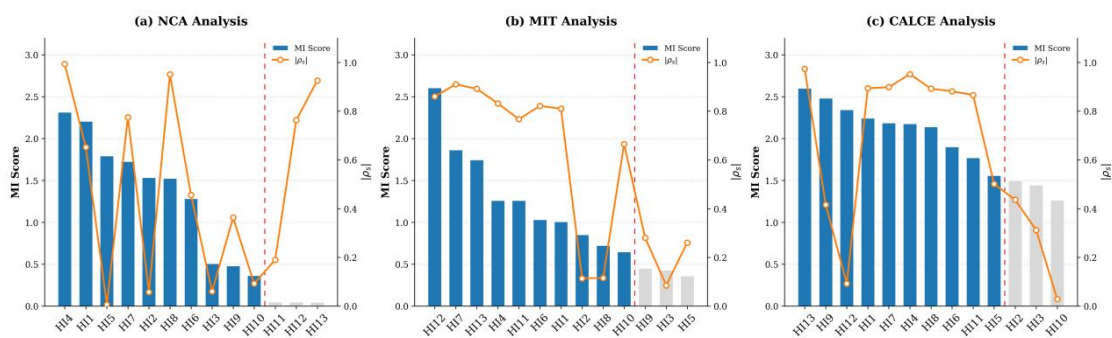

Fig. 2. MI scores and Spearman correlation coefficients ($|\rho_s|$) for the 13 candidate HIs across three datasets: (a) NCA, (b) MIT, and (c) CALCE. Blue bars represent MI scores and orange lines denote $|\rho_s|$ values. The dashed vertical line indicates the selection threshold, with grey bars corresponding to discarded features.

### 3.3 Datasets and Partition Strategy

To evaluate the generalization capability of the proposed framework across heterogeneous electrochemical systems, three representative datasets covering mainstream lithium-ion battery chemistries—namely NCA, LFP, and LCO—are employed. These datasets exhibit distinct degradation characteristics in terms of aging rate, trajectory smoothness, and inter-cell variability, providing a comprehensive benchmark for model evaluation. The corresponding SOH evolution trajectories are shown in Fig. 3.

**NCA Dataset:** Released by Tongji University, this dataset consists of commercial high-energy-density $LiNi_{0.86}Co_{0.11}Al_{0.03}O_2$ pouch cells cycled under accelerated aging conditions, including an ambient temperature of 45°C and a 5C high-rate charging protocol. Twenty-eight full-life cells are selected. These cells exhibit rapid and relatively monotonic degradation with limited inter-cell variability, making the dataset suitable for evaluating model performance under severe thermal and kinetic stress conditions [33].

**MIT Dataset:** Provided by the Massachusetts Institute of Technology, this dataset comprises $LiFePO_4$ 18650 cells with long cycle life and relatively smooth degradation trajectories. To ensure data quality and physical consistency, a systematic anomaly filtering procedure is applied. Specifically, cells with complete data loss due to equipment failure (Channels 4 and 8), samples that do not reach the 80% EOL threshold (Channels 13, 19, 21, 22, 31), and cells exhibiting non-physical capacity recovery after prolonged rest periods (Channels 1–3, 5, 6) are excluded. After filtering, 34 cells are retained, yielding a dataset suitable for evaluating long-horizon degradation modeling with a relatively large sample size [34].

**CALCE Dataset:** Sourced from the University of Maryland, this dataset comprises $LiCoO_2$ 18650 cells commonly used in consumer electronics. Four cells from the CS2 series are selected. Compared with other datasets, these cells exhibits stronger cell-to-cell variability and occasional capacity regeneration behavior, making it suitable for assessing model robustness under irregular and non-monotonic degradation patterns [35].

Data partitioning is performed at the battery-cell level rather than at the cycle level, ensuring that all cycles from a given cell are assigned exclusively to either the training or test set. This strategy prevents information leakage caused by temporal overlap and ensures that the evaluation reflects true cross-cell generalization rather than intra-cell interpolation [22]. For datasets with sufficient sample size (NCA and MIT), a fixed training-test split is adopted. For the smaller CALCE dataset, leave-one-out cross-validation (LOOCV) is employed to reduce bias arising from limited sample diversity. In all cases, normalization parameters and feature selection procedures are computed exclusively from the training partition and then applied to the test set, maintaining strict separation between training and evaluation.

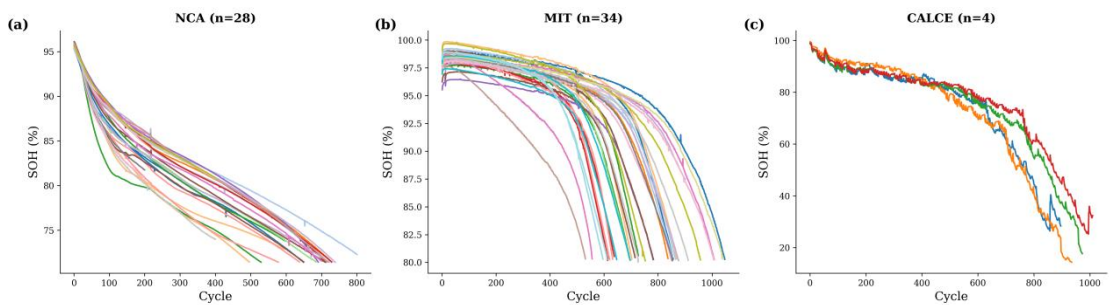

Fig. 3. SOH evolution trajectories of the three benchmark datasets: (a) NCA, (b) MIT, and (c) CALCE. Each curve represents an individual battery cell, illustrating differences in degradation rate, cycle life, and inter-cell variability across electrode chemistries.

## 3.4 Quantum-Enhanced Recurrent Modeling Framework

Recurrent neural networks are well suited for modeling battery degradation because of their ability to capture long-horizon temporal dependencies. Among them, LSTM networks are particularly effective in mitigating gradient vanishing through gated memory mechanisms, which makes them a natural choice for SOH prediction [27]. In classical LSTM, state transitions are governed by affine transformations followed by element-wise activation functions. Although effective in many applications, this formulation may limit the model's ability to capture structured feature interactions arising from coupled electrochemical degradation processes.

To address this limitation, we introduce a quantum-enhanced recurrent modeling framework in which the affine transformations in LSTM gating mechanisms are replaced by VQCs. This design preserves the original memory update structure while enabling more flexible nonlinear transformations within each gating unit.

From a functional perspective, the proposed QLSTM defines a quantum-enhanced recurrent mapping in which classical linear projections are replaced by parameterized quantum transformations operating in a high-dimensional quantum state space. This formulation provides an alternative mechanism for modeling nonlinear feature interactions while preserving the temporal modeling capability of recurrent networks. Importantly, the overall structure of the QLSTM cell remains consistent with that of the classical LSTM, ensuring compatibility with standard sequence modeling workflows. As a result, the proposed framework enhances representational capacity without altering the underlying recurrent learning paradigm.

**3.4.1 Quantum-Enhanced Gating Mechanism**

As illustrated in Fig. 4, the recurrent structure and memory update mechanism of LSTM are preserved, while the affine transformations within each gate are replaced by parameterized VQC modules. Given the input $x_t$ and the previous hidden state $h_{t-1}$, we form a concatenated vector $v_t=[h_{t-1},x_t]$. This vector serves as the input to the gating units. The gate activations are defined as:

$$f_t = \sigma(VQC_f(\mathbf{v}_t)) \tag{4}$$

$$i_t = \sigma(VQC_i(\mathbf{v}_t)) \tag{5}$$

$$\tilde{c}_t = \tanh(VQC_c(\mathbf{v}_t)) \tag{6}$$

$$c_t = f_t \odot c_{t-1} + i_t \odot \tilde{c}_t \tag{7}$$

$$o_t = \sigma(VQC_o(v_t)) \tag{8}$$

$$h_t = o_t \odot \tanh(c_t) \tag{9}$$

where $f_t$, $i_t$, $\tilde{c}_t$, and $o_t$ denote the forget gate, input gate, candidate memory state, and output gate, respectively.

In this formulation, each VQC performs a parameterized nonlinear transformation through unitary evolution followed by measurement. Compared to classical affine mappings, this mechanism enables richer feature interactions among input components. By embedding VQC-based transformations into the gating mechanism, the QLSTM retains the temporal memory properties of LSTM while providing additional flexibility for modeling complex degradation dynamics.

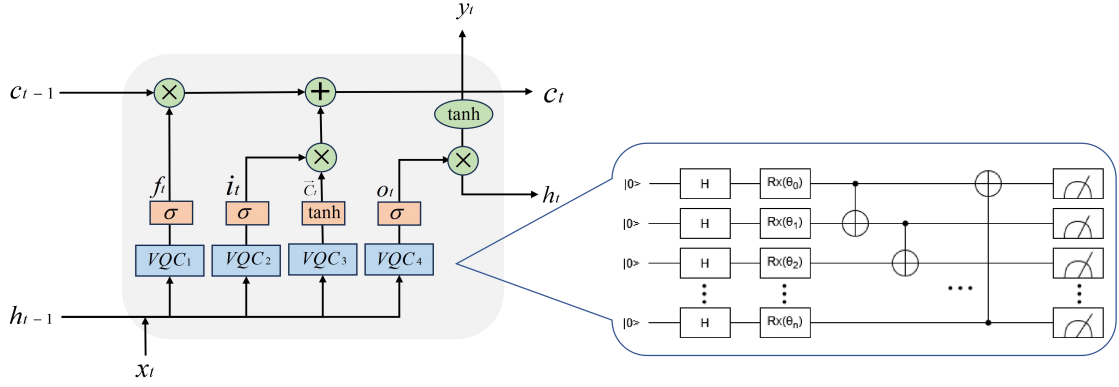

Fig. 4. Schematic of the proposed QLSTM cell.

**3.4.2 Variational Quantum Circuit Architecture**

Each VQC consists of three functional stages: encoding, variational evolution, and measurement, as illustrated in Fig. 5.

Encoding Layer: The input vector $v_t$ is normalized and embedded into the quantum state space through angle encoding [36]. Each qubit is initialized into a superposition state via a Hadamard gate, followed by parameterized rotation gates Ry and Rz with rotation angles determined by the input features. This process encodes classical information into the quantum state representation.

Variational Layer: A hardware-efficient ansatz is adopted as the learnable component of the circuit [37]. Entanglement among qubits is introduced via controlled-NOT (CNOT) gates, enabling interactions between encoded features. Parameterized single-qubit rotation gates R($\alpha$, $\beta$, $\gamma$) are then applied, with parameters optimized during training [38]. This layer functions as a nonlinear feature approximator, adaptively modeling coupled degradation dynamics that classical affine transformations often fail to capture.

Measurement Layer: The evolved quantum state is projected back into the classical domain by measuring the expectation value of the Pauli-Z operator on each qubit, resulting in a real-valued output vector in the range [−1,1] [39]. These outputs are then used directly as gating signals to regulate recurrent state updates within the QLSTM cell.

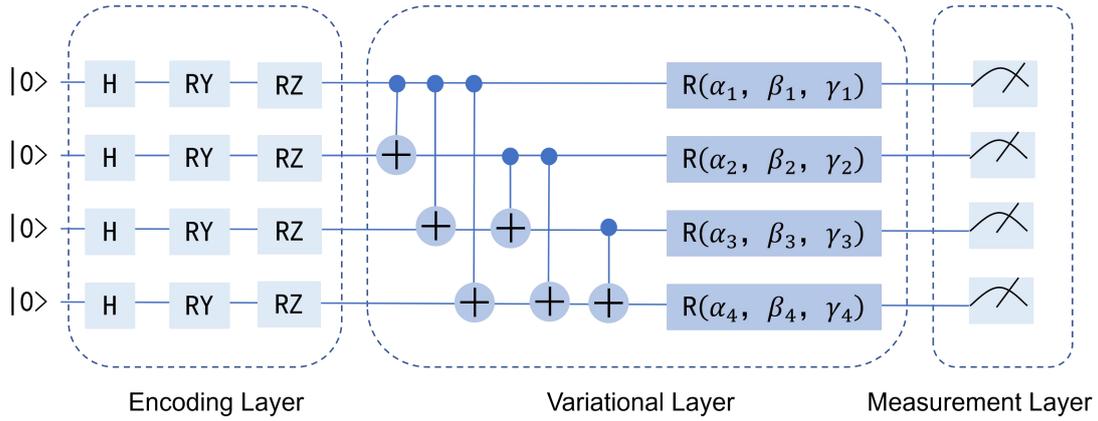

Fig. 5. Circuit architecture of the proposed VQC module.

Overall, this shallow circuit design enables expressive nonlinear transformations while maintaining a compact parameter space. Its limited depth ensures compatibility with gradient-based optimization and NISQ hardware, making the circuit suitable for hybrid quantum–classical training and practical quantum simulations.

### 3.5 Training and Computational Considerations

The proposed QLSTM model is trained within a hybrid quantum–classical optimization framework, in which classical network parameters and VQC parameters are jointly optimized in a unified training loop. Gradients for the classical components are computed using standard backpropagation, while gradients for the quantum components are obtained via the parameter-shift rule, which enables analytic gradient evaluation of expectation-based measurements. Specifically, for a variational parameter θ associated with a single-qubit rotation gate, the gradient of an expectation value f(θ) is computed as:

$$\frac{\partial f(\theta)}{\partial \theta} = \frac{f\left(\theta+\frac{\pi}{2}\right) - f\left(\theta-\frac{\pi}{2}\right)}{2} \quad (10)$$

This formulation requires two forward evaluations of the quantum circuit for each parameter and avoids the numerical inaccuracies associated with finite-difference approximations. Gradients from both the classical and quantum components are integrated through an automatic differentiation pipeline and used to update all model parameters via the Adam optimizer [40]. This design enables coordinated optimization of classical and quantum components within a single training process.

To ensure stable training, all input features are normalized and fixed-length sequence windows are constructed. Gradient clipping is applied to mitigate instability during recurrent updates, and a learning rate scheduler is used to gradually decrease the learning rate over training epochs to facilitate convergence.

All experiments are conducted in a hybrid simulation environment, where quantum circuits are executed on a simulator and integrated with classical deep learning frameworks. This setup allows efficient experimentation while maintaining compatibility with near-term quantum hardware.

The computational cost of VQC gradient evaluation grows linearly with the number of variational parameters when using the parameter-shift rule, assuming fixed circuit depth and measurement repetitions. In practice, this overhead is mitigated by the use of shallow circuit designs and efficient integration with classical network components. Overall, the proposed training scheme balances the expressive capability of quantum-enhanced transformations with the stability and efficiency of classical optimization, enabling effective sequence modeling under constrained quantum resources.

## 4. Experimental Results and Discussion

This section presents a systematic evaluation of the proposed QLSTM framework from multiple perspectives, including predictive accuracy, cross-chemistry generalization, structural contributions, and robustness under realistic quantum noise. Section 4.1 describes the experimental setup. Section 4.2 introduces the evaluation metrics. Section 4.3 reports predictive performance across datasets, followed by a structural ablation study in Section 4.4. Section 4.5 analyzes qubit scaling behavior, and Section 4.6 evaluates robustness under simulated NISQ noise. Section 4.7 concludes with a discussion of the main findings.

### 4.1 Experimental Setup

All experiments were conducted on a hybrid quantum-classical computing platform developed by Yiwei Quantum Technology Co., Ltd. The classical components were implemented in PyTorch, and the quantum circuits were simulated using TorchQuantum. All computations were performed on a CentOS 8.5 server equipped with an NVIDIA A100 GPU (80 GB) and 512 GB RAM.

The proposed QLSTM adopts the variational quantum circuit architecture described in Section 3. Unless otherwise specified, a 4-qubit configuration with shallow circuit depth is employed to balance representational capacity and computational efficiency under NISQ constraints. Measurement outputs, obtained as expectation values of Pauli-Z operators, are integrated into the recurrent gating mechanism.

Model optimization was performed using the Adam optimizer with mean squared error (MSE) as the loss function. Hyperparameters such as learning rate, batch size, and dropout rate were tuned for each dataset to accommodate the heterogeneous degradation characteristics of different battery chemistries, as summarized in Table 2. All models were trained for 100 epochs without early stopping to ensure a consistent training budget. To ensure fair comparison, all models were trained using identical data partitions, input features, normalization procedures, and optimization pipelines. Hyperparameters for baseline models were tuned within comparable search ranges to avoid bias toward any specific architecture.

Baseline models include CNN, GRU, and classical LSTM with 128 hidden units, all implemented under the same training framework as QLSTM. Each experiment was repeated three times with different random seeds, and mean performance values with standard deviations are reported.

Table 2. Hyperparameter settings of the proposed QLSTM for the three battery datasets.

| Dataset | Layers | Hidden Units | Learning Rate | Batch Size | Dropout |
|---|---|---|---|---|---|
| MIT | 1 | 128 | 0.001 | 64 | 0 |
| NCA | 1 | 128 | 0.001 | 64 | 0 |
| CALCE | 1 | 128 | 0.01 | 64 | 0.2 |

**4.2 Evaluation Metrics**

Model performance is evaluated using three complementary regression metrics: Mean Absolute Error (MAE), Root Mean Square Error (RMSE), and the coefficient of determination ($R^2$). These metrics collectively quantify point-wise prediction accuracy, sensitivity to large deviations, and trajectory-level fitting quality.

The MAE measures the average absolute deviation between predicted and true SOH values:

$$MAE = \frac{1}{n}\sum_{i=1}^{n}|y_i - \hat{y}_i| \tag{11}$$

The RMSE places a heavier penalty on larger errors and is particularly informative for long-horizon degradation modeling, where substantial deviations may indicate a failure to capture abrupt aging transitions:

$$RMSE = \sqrt{\frac{1}{n}\sum_{i=1}^{n}(y_i - \hat{y}_i)^2} \tag{12}$$

The coefficient of determination $R^2$ is computed as:

$$R^2 = 1 - \frac{\sum_{i=1}^{n}(y_i - \hat{y}_i)^2}{\sum_{i=1}^{n}(y_i - \bar{y}_i)^2} \tag{13}$$

where $y_i$ and $\hat{y}_i$ are the true and predicted SOH values at cycle i, $\bar{y}_i$ is the mean of the true SOH values, and n is the number of test samples. Values of $R^2$ closer to 1 indicate better reconstruction of the degradation trajectory. These metrics are well suited to long-horizon SOH prediction because they jointly reflect local accuracy and global consistency, both of which are critical for reliable battery health prognostics.

**4.3 Predictive Performance Across Chemistries**

Following the cell-level partition strategy described in Section 3.3, all test cells were completely unseen during training. For the MIT dataset, 26 of 34 cells were used for training and the remaining 8 for testing. For the NCA dataset, 22 of 28 cells were allocated to training and 6 to testing. For the CALCE dataset, a LOOCV protocol was applied across the four benchmark cells (CS2_35 to CS2_38), with one cell held out per fold and results averaged across folds. In all cases, normalization parameters were computed exclusively from the training set to avoid data leakage.

**Overall Accuracy Across Chemistries.** As summarized in Table 3 and Fig. 6, QLSTM consistently achieves the best performance across all datasets and evaluation metrics. On the CALCE dataset, QLSTM attains an MAE of $0.0112 \pm 0.0014$, which is approximately 20% lower than that of classical LSTM ($0.0140 \pm 0.0015$), while also achieving the lowest RMSE ($0.0148 \pm 0.0016$). On the NCA dataset, QLSTM achieves an RMSE of $0.0091 \pm 0.0035$ and $R^2$ of $0.9965 \pm 0.0024$, substantially outperforming CNN ($0.0314 \pm 0.0063$) and GRU ($0.0233 \pm 0.0088$) in terms of RMSE. On the MIT dataset, QLSTM reaches an MAE of $0.0011 \pm 0.0003$ and an $R^2$ of $0.9977 \pm 0.0024$, indicating strong capability in modeling long-horizon temporal dependencies. Although GRU performs comparably to, or slightly better than, classical LSTM on the MIT dataset in certain metrics, QLSTM consistently outperforms both, indicating that its advantage extends beyond LSTM to recurrent architectures more broadly.

**Cross-Chemistry Generalization.** QLSTM maintains robust performance across LFP (MIT), LCO (CALCE), and NCA chemistries, which differ in smoothness, nonlinearity, and regeneration behavior. As shown in Fig. 6, QLSTM accurately tracks SOH trajectories for all chemistries, suggesting that the learned representation captures degradation patterns rather than chemistry-specific features. This result highlights the cross-domain robustness of the framework.

**Stability and Variance Reduction.** In addition to improved predictive accuracy, QLSTM also exhibits lower variance across repeated runs. For example, on the MIT dataset, the MAE standard deviation of QLSTM (±0.0003) is lower than that of both LSTM (±0.0005) and CNN (±0.0005), with similar trends observed across other datasets. Although performance variance increases on the NCA dataset because of stronger data heterogeneity, QLSTM remains competitive with, or more stable than, the baseline models. This reduced variance indicates more consistent convergence behavior and lower sensitivity to random initialization, both of which are particularly important in long-horizon SOH prediction, where prediction errors may accumulate over time.

One possible contributing factor is the structural constraint imposed by unitary transformations within VQC-based gates. Compared with unconstrained affine mappings, the compact parameterization of VQCs may act as an implicit regularization, reducing sensitivity to stochastic perturbations and promoting stable learning dynamics. A rigorous theoretical analysis of this effect, however, is beyond the scope of this work.

**Mechanistic Interpretation.** The observed performance gains can be attributed to the modified recurrent state-transition mechanism in QLSTM. By embedding structured unitary transformations directly into gating functions, QLSTM introduces richer nonlinear feature interactions within the recurrent update process. In contrast, classical LSTM and GRU rely on affine transformations followed by element-wise nonlinearities, which may be less effective in capturing strongly coupled degradation dynamics.

This advantage is particularly evident in datasets with pronounced nonlinearity and inter-cell variability, such as CALCE and NCA. As further confirmed by the ablation study in Section 4.4, the primary source of performance improvement arises from quantum-enhanced gating rather than input-level transformations.

Overall, these results suggest that QLSTM provides a more expressive yet structurally constrained recurrent representation for modeling complex battery degradation processes.

Table 3. Comparative performance of SOH prediction across CALCE, NCA, and MIT datasets (mean ± standard deviation). Best results are highlighted in bold.

| Dataset | Model | Mean MAE | Mean RMSE | Mean $R^2$ |
|---|---|---|---|---|
| CALCE | QLSTM | **0.0112 ± 0.0014** | **0.0148 ± 0.0016** | **0.9945 ± 0.0018** |
| | LSTM | 0.0140 ± 0.0015 | 0.0191 ± 0.0035 | 0.9924 ± 0.0019 |
| | GRU | 0.0143 ± 0.0046 | 0.0187 ± 0.0048 | 0.9929 ± 0.0021 |
| | CNN | 0.0178 ± 0.0025 | 0.0212 ± 0.0019 | 0.9900 ± 0.0034 |
| NCA | QLSTM | **0.0111 ± 0.0040** | **0.0091 ± 0.0035** | **0.9965 ± 0.0024** |
| | LSTM | 0.0141 ± 0.0049 | 0.0128 ± 0.0047 | 0.9939 ± 0.0037 |
| | GRU | 0.0205 ± 0.0083 | 0.0233 ± 0.0088 | 0.9830 ± 0.0099 |
| | CNN | 0.0299 ± 0.0064 | 0.0314 ± 0.0063 | 0.9727 ± 0.0095 |
| MIT | QLSTM | **0.0011 ± 0.0003** | **0.0020 ± 0.0010** | **0.9977 ± 0.0024** |
| | LSTM | 0.0015 ± 0.0005 | 0.0028 ± 0.0014 | 0.9959 ± 0.0040 |
| | GRU | 0.0014 ± 0.0004 | 0.0024 ± 0.0012 | 0.9966 ± 0.0039 |
| | CNN | 0.0014 ± 0.0005 | 0.0032 ± 0.0021 | 0.9948 ± 0.0052 |

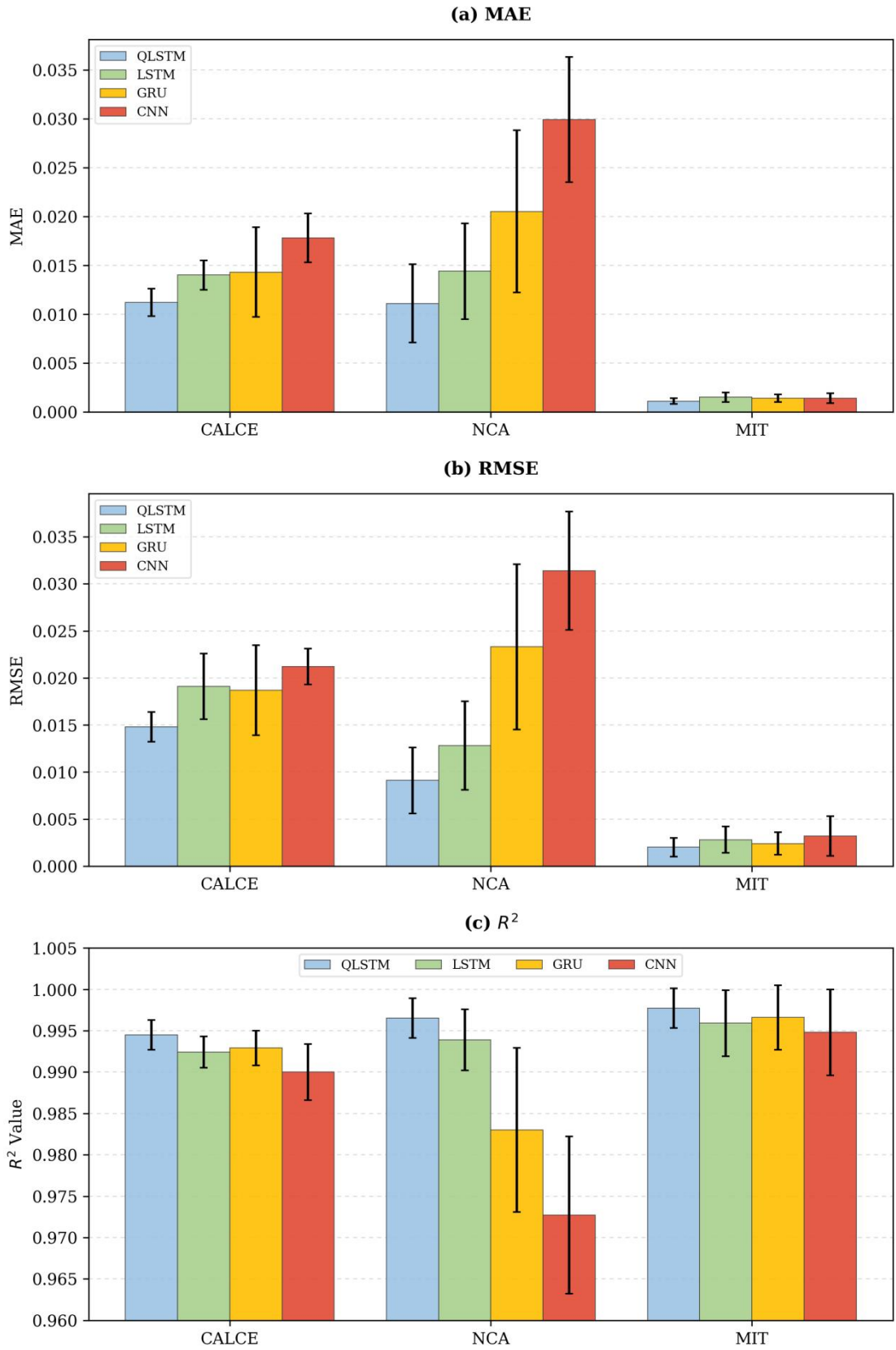

Fig. 6. Performance comparison of QLSTM and classical baseline models across the three datasets: (a) MAE, (b) RMSE, and (c) $R^2$. Error bars denote one standard deviation across repeated trials.

**4.4 Structural Ablation Study**

To further investigate the source of performance gains observed in Section 4.3, a structural ablation study was conducted on the CALCE dataset, which is characterized by pronounced inter-cell variability and nonlinear degradation behavior. Four architectural variants were evaluated under identical training configurations, dataset partitions, and optimization protocols, thereby enabling a controlled comparison of the contributions of individual model components. The evaluated variants are defined as follows:

Classical LSTM: A standard LSTM with affine gate transformations, used as the baseline model.

NG-LSTM (Nonlinear-Gated LSTM): An LSTM variant in which each affine gate is replaced by a shallow multilayer perceptron consisting of a linear layer, layer normalization, a GELU activation, and a second linear layer (Linear-LayerNorm-GELU-Linear), thereby introducing classical nonlinearity into the gating mechanism.

QE-LSTM (Quantum-Embedding LSTM): A variant in which a VQC is applied as an external quantum feature transformation before recurrent processing, while the internal gating mechanism remains classically affine.

Proposed QLSTM: The full model, in which VQC-based transformations are integrated directly into all four internal gating units, fully replacing classical affine gate transformations.

This progression disentangles three distinct factors: (i) classical nonlinearity in gating (NG-LSTM), (ii) quantum feature embedding (QE-LSTM), and (iii) quantum-enhanced gating for recurrent state transitions (QLSTM). The ablation framework therefore provides a structured basis for identifying the principal source of performance improvement.

**Quantitative Results.** The LOOCV results summarized in Table 4 provide consistent evidence that QLSTM achieves the best performance across all four CALCE cells, with the lowest MAE and RMSE and $R^2$ values above 0.99 in every fold. For example, on CS2_36, QLSTM reduces the RMSE from 0.0242 for LSTM to 0.0153, corresponding to an improvement of approximately 36.8%. Similar trends are observed across all cells, demonstrating that the performance gain is both consistent and robust under diverse degradation patterns.

QE-LSTM outperforms classical LSTM in every fold, suggesting that quantum feature embedding improves representational quality even when the gating mechanism remains classical. Notably, QE-LSTM and QLSTM share the same quantum circuit design and parameterization, differing only in whether quantum transformations are applied at the input level or within the gating units. By contrast, NG-LSTM shows greater variability: it outperforms QE-LSTM on some cells (e.g., CS2_36 and CS2_37) but underperforms on others, including CS2_35 and CS2_38.

This inconsistency suggests that introducing classical nonlinearity into the gating mechanism alone is insufficient to ensure reliable generalization under limited training data and heterogeneous degradation behavior.

**Mechanism-Level Interpretation.** To better understand the observed performance differences, three pairwise comparisons are considered. First, comparing LSTM and NG-LSTM shows that introducing classical nonlinearity into gating produces inconsistent benefits, suggesting that increased functional complexity alone does not ensure improved sequence modeling performance. Second, the comparison between LSTM and QE-LSTM indicates that quantum feature embedding enhances input representations; however, its impact is limited when the recurrent gating mechanism itself remains classically affine. Third, and most importantly, the comparison between QE-LSTM and QLSTM highlights that embedding VQC directly into the gating units produces substantial and consistent improvements.

These observations suggest that the key advantage of QLSTM lies in enhancing the recurrent gating mechanism itself. By introducing structured unitary transformations at each recurrent step, QLSTM enables richer and more stable recursive dynamics, rather than relying solely on enhanced input representations. Compared with classical affine mappings, this structured evolution provides a more expressive yet constrained progression of hidden states.

A further comparison between NG-LSTM and QLSTM reinforces this conclusion. Although both models introduce additional nonlinearity into the gating mechanism, only QLSTM achieves consistent improvements across all folds. This indicates that the observed gains cannot be attributed to nonlinearity alone, but are more likely associated with the geometric and structural properties of quantum unitary transformations.

**Implications.** Battery degradation is inherently cumulative, and its long-term dependencies emerge through repeated state transitions. The ablation results indicate that improvements limited to input representation are insufficient when the recurrent update mechanism remains based on classical affine gating transformations, and that classical nonlinearity alone does not reliably capture the coupled degradation dynamics present in heterogeneous battery datasets. These results therefore indicate that quantum-enhanced gating, rather than feature-level transformation, is the primary driver of performance improvement. Although these conclusions remain empirical, they further suggest that embedding structured transformations within recurrent operators is a promising direction for sequence modeling in complex physical systems.

Table 4. Results of the ablation study on the CALCE dataset using LOOCV.

|  | Model | MAE | RMSE | $R^2$ |
|---|---|---|---|---|
| CS2_35 | QLSTM | **0.011** | **0.0146** | **0.9938** |
|  | QE-LSTM | 0.0135 | 0.0173 | 0.9912 |
|  | NG-LSTM | 0.015 | 0.0191 | 0.9894 |
|  | LSTM | 0.0135 | 0.018 | 0.9905 |
| CS2_36 | QLSTM | **0.0118** | **0.0153** | **0.994** |
|  | QE-LSTM | 0.0146 | 0.0188 | 0.9932 |
|  | NG-LSTM | 0.0139 | 0.0174 | 0.9935 |
|  | LSTM | 0.0162 | 0.0242 | 0.9918 |
| CS2_37 | QLSTM | **0.0094** | **0.0127** | **0.9971** |
|  | QE-LSTM | 0.0132 | 0.0161 | 0.9953 |
|  | NG-LSTM | 0.0127 | 0.0157 | 0.9956 |
|  | LSTM | 0.0134 | 0.0166 | 0.9951 |
| CS2_38 | QLSTM | **0.0127** | **0.0166** | **0.9929** |
|  | QE-LSTM | 0.0134 | 0.0167 | 0.9928 |
|  | NG-LSTM | 0.0130 | 0.0180 | 0.9917 |
|  | LSTM | 0.0129 | 0.0174 | 0.9923 |

**4.5 Qubit-Scale Analysis**

To examine the effect of quantum circuit capacity on predictive performance, we conducted a qubit-scale analysis on the CALCE dataset. The number of qubits was varied as n = 4, 6, 8, 10, and 12, while all other training configurations were kept unchanged to isolate the effect of circuit width. In the proposed architecture, each VQC module receives a concatenated vector composed of the selected input features and the previous hidden state, which is then encoded into an n-qubit quantum state through angle encoding. Increasing the number of qubits expands the Hilbert-space dimension exponentially, potentially enabling richer feature interactions within the gating mechanism.

As illustrated in Fig. 7, the relationship between qubit count and prediction error is not strictly monotonic. Although larger configurations (n = 10 or 12) generally achieve competitive or improved performance, intermediate configurations such as n = 8 may exhibit reduced accuracy. This non-monotonic behavior is consistently observed across all test cells, indicating that increasing quantum capacity alone does not guarantee improved predictive performance.

The marginal performance gain obtained by increasing n from 4 to 12 remains modest across all cells. At the same time, higher qubit counts increase circuit depth and parameter complexity, thereby introducing additional computational overhead. Under identical training configurations, this trade-off highlights that performance improvement is not proportional to circuit size. For this reason, a 4-qubit configuration is adopted as the default setting in this study, as it provides a practical balance between predictive performance and computational efficiency under NISQ hardware constraints.

The observed non-monotonic trend suggests that the effectiveness of quantum circuit scaling depends not only on the size of the Hilbert space, but also on how efficiently the added capacity can be utilized. Under a fixed circuit depth and encoding strategy, adding qubits introduces more degrees of freedom, which may lead to optimization difficulty, inefficient parameter usage, or increased sensitivity to initialization. These effects may be related to known training challenges in variational quantum circuits, where increased system size can lead to flatter optimization landscapes or reduced trainability. In addition, the scaling behavior varies across cells, reflecting the heterogeneous nature of battery degradation dynamics. This indicates that the optimal circuit size is data-dependent rather than universal.

Taken together, these results highlight that quantum circuit scaling in recurrent architectures is governed by a nontrivial interplay among representational capacity, trainability, and data-dependent structure. Overall, these findings suggest that effective capacity utilization, rather than raw Hilbert-space expansion, is the key factor in achieving performance gains. Identifying principled scaling strategies for quantum-enhanced sequence models therefore remains an important direction for future research.

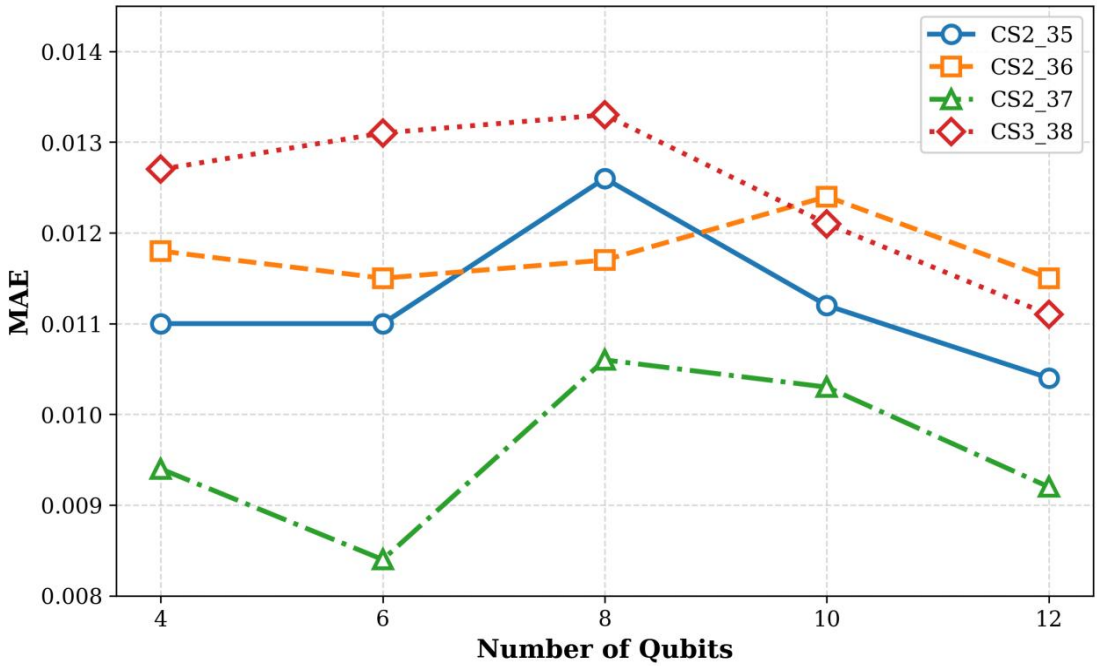

Fig. 7. MAE as a function of qubit count (n∈ {4,6,8,10,12}) for the four CALCE test cells.

**4.6 Robustness under Simulated NISQ Noise**

To assess the robustness of QLSTM under realistic quantum hardware conditions, experiments were conducted using a simulated noise model representative of NISQ devices. Bit-flip noise with probabilities p∈ {0.01,0.02,0.05} was injected into the VQC modules, while all other training and evaluation settings were kept unchanged.

As shown in Fig. 8, predictive performance degrades gradually as the noise level increases, without any abrupt collapse across datasets. This smooth degradation trend is consistently observed across all datasets, indicating stable model behavior under increasing noise levels. On the MIT dataset, $R^2$ decreases from 0.9977 in the noiseless setting to 0.9957 at $p = 0.05$, while MAE increases from 0.0011 to 0.0021. Similar trends are observed for the NCA and CALCE datasets, where $R^2$ continues to exceed 0.99 even at the highest noise level.

The gradual nature of the performance degradation indicates that the proposed architecture retains a degree of robustness under moderate quantum noise. One possible reason is that the model outputs are derived from expectation values rather than single-shot measurements. Because expectation values reflect statistical averages over quantum states, localized bit-flip errors may introduce bounded perturbations to the gate activations, thereby limiting their influence on the final predictions. This averaging effect may help mitigate the impact of stochastic noise at the measurement level. In addition, the shallow circuit depth and compact qubit configuration reduce cumulative noise effects relative to deeper quantum circuits.

From a modeling perspective, embedding VQC within the gating units localizes quantum computation to individual recurrent steps, which may help prevent excessive error accumulation over long temporal sequences. This localized integration contrasts with deeper quantum architectures where noise may accumulate across layers, thereby contributing to improved stability. This structural property likely contributes to the observed stability under noise, although a rigorous theoretical analysis is beyond the scope of this study.

Practically, the circuit design is consistent with the capabilities of current NISQ hardware. The use of a shallow, hardware-efficient ansatz and a small number of qubits (e.g., 4 qubits in the main experiments) ensures compatibility with the coherence times and gate fidelities of contemporary superconducting and trapped-ion platforms. These design choices align the proposed architecture with near-term hardware constraints.

It should be noted that these evaluations were performed under classical simulation with a simplified noise model and therefore do not capture all sources of hardware noise. In particular, correlated noise, decoherence effects, and gate imperfections are not fully modeled in this study. Nevertheless, the observed moderate sensitivity suggests that small-scale quantum-enhanced gating can operate reliably under realistic conditions.

Overall, these results demonstrate that the QLSTM framework achieves a favorable balance between predictive accuracy and noise robustness, supporting its potential feasibility for near-term quantum-enhanced sequence modeling applications. Taken together, these findings suggest that the observed robustness is attributable, at least in part, to the quantum-enhanced gating mechanism rather than to a specific experimental configuration.

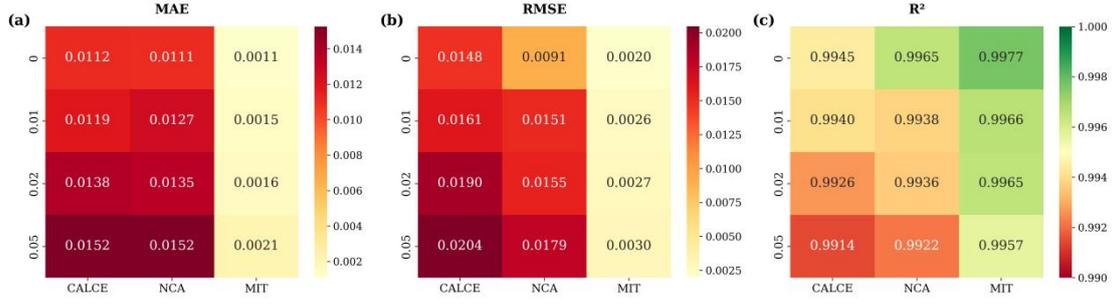

Fig. 8. Performance degradation of QLSTM under simulated bit-flip noise across three datasets: (a) MAE, (b) RMSE, and (c) $R^2$. Rows correspond to noise probabilities $p \in \{0, 0.01, 0.02, 0.05\}$.

**4.7 Discussion**

Compared with approaches that use quantum components primarily as external feature transformations, the present framework integrates quantum computation directly into the recurrent gating mechanism responsible for state transitions. The experimental results presented in Sections 4.3 – 4.6 provide a comprehensive evaluation of QLSTM in terms of predictive accuracy, structural contributions, capacity scaling, and robustness under realistic noise conditions. Overall, these findings consistently indicate that embedding VQCs within recurrent gating mechanisms provides a structurally grounded enhancement over classical sequence models. This shift from feature-level transformation to gating-level integration represents a fundamental change in how quantum computation is incorporated into sequence modeling.

The ablation study in Section 4.4 shows that the primary source of performance gain stems from enhancing the recurrent gating mechanism responsible for state transitions rather than solely improving input representations. Embedding structured transformations directly into the recurrent update process appears more effective than increasing functional complexity at the input level. Notably, the comparison between QE-LSTM and QLSTM isolates the gating mechanism as the only structural difference, providing strong evidence that quantum-enhanced gating is the dominant factor driving performance improvement. The structural constraints imposed by quantum transformations may also contribute to improved generalization and training stability by acting as a form of implicit regularization.

The qubit-scale analysis in Section 4.5 further reveals that increasing quantum capacity does not lead to monotonic performance improvement. Instead, predictive performance depends on a nontrivial interplay among representational capacity, trainability, and data-dependent structure. This finding suggests that, for quantum-enhanced sequence models, effective use of representational capacity may be more important than raw Hilbert space dimensionality, thereby underscoring the need for principled scaling strategies. In particular, these results indicate that circuit scalability should be evaluated in terms of trainable capacity rather than nominal Hilbert-space size.

The robustness analysis in Section 4.6 demonstrates that the proposed architecture maintains stable performance under moderate noise, supporting its potential feasibility for near-term quantum hardware. The combination of shallow circuit design, expectation-based measurement, and localized quantum operations within gating units appears to mitigate noise effects, although further validation on real quantum devices remains necessary. This suggests that robustness is closely related to architectural design choices, rather than being solely dependent on hardware quality.

Collectively, the results support the use of quantum-enhanced recurrent architectures as a promising approach for modeling complex temporal systems with strong nonlinearity and long-term dependencies. More broadly, the proposed framework highlights the importance of embedding quantum operations within dynamical operators, rather than treating them as static feature transformations, in hybrid quantum–classical learning systems. Beyond battery health prediction, the framework may be applicable to a broader class of sequence modeling problems in physical and engineering systems, where both structured transformations and recursive dynamics are essential.

Several open questions remain. First, a deeper theoretical understanding of the expressiveness and generalization behavior of quantum-enhanced gating mechanisms is still needed. Second, the interplay between circuit design, encoding strategies, and trainability warrants further investigation. Third, extending the framework to larger-scale quantum hardware and exploring more efficient hybrid classical-quantum training schemes may improve its practical applicability. In particular, establishing theoretical links between quantum circuit structure and recurrent dynamics remains an important open problem.

These directions point to a broader research agenda at the intersection of quantum computing and sequence modeling, where architectural design, physical constraints, and learning dynamics must be considered jointly. Future progress in this area is likely to depend on co-design across algorithms, models, and quantum hardware.

## 5. Conclusion

This paper presents a quantum-enhanced recurrent framework, termed QLSTM, for battery SOH prediction, in which variational quantum circuits are embedded directly into the gating mechanisms of LSTM. By replacing classical affine transformations with parameterized unitary operations, the proposed model introduces structured nonlinear transformations into the recurrent state-transition process. This design enables quantum computation to act directly on the dynamical evolution of hidden states, rather than serving as an external feature transformation module.

Extensive experiments on three benchmark battery datasets demonstrate that QLSTM achieves consistently improved predictive accuracy and reduced variance compared to classical sequence models. Systematic ablation studies reveal that the primary performance gains arise from modifying the recurrent operator itself rather than enhancing input representations. In particular, controlled comparisons confirm that quantum-enhanced gating, rather than classical nonlinearity or input-level quantum embedding, is the dominant factor driving performance improvements. Additional analyses on qubit scaling and noise robustness indicate that model performance is governed by a balance between representational capacity and trainability, and that the proposed architecture maintains stable behavior under simulated quantum noise. These results further suggest that effective capacity utilization, rather than raw Hilbert-space expansion, is critical for quantum-enhanced sequence models.

Overall, this work provides empirical evidence that integrating quantum computational primitives into recurrent neural architectures can enhance sequence modeling capability. More importantly, it points to a design paradigm in which quantum operations are embedded within recurrent gating mechanisms for state transitions, offering a structurally grounded approach to modeling complex temporal dynamics beyond the representational scope of classical affine transformations.

Future work will focus on theoretical analysis of the expressiveness and generalization properties of quantum-enhanced recurrent models, validation on real quantum hardware, and extension to broader classes of temporal prediction tasks. In particular, establishing theoretical connections between quantum circuit structures and recurrent dynamical behavior remains an important direction for future research.

**Acknowledgements**

The authors acknowledge Yiwei Quantum Technology Co., Ltd. (Hefei 230088, China) for providing the hybrid quantum-classical computing platform used in this study.

**Conflict of Interest**

The authors declare no conflict of interest.

**Code Availability Statement**

The code that supports the findings of this study is available from the corresponding author upon reasonable request.

**References**

[1] J. Vetter, P. Novak, M. R. Wagner et al., "Ageing mechanisms in lithium-ion batteries," *J. Power Sources*, vol. 147, no. 1-2, pp. 269-281, 2005.

[2] S. Hochreiter and J. Schmidhuber, "Long short-term memory," *Neural Comput.*, vol. 9, no. 8, pp. 1735-1780, Nov. 1997.

[3] D. Roman, S. Saxena, V. Robu et al., "Machine learning pipeline for battery state-of-health estimation," *Nat. Mach. Intell.*, vol. 3, no. 5, pp. 447-456, 2021.

[4] K. L. Soon and L. T. Soon, "A hybrid quantum neural network and classical gated recurrent unit for battery state of health forecasting incorporating SHAP analysis," *J. Energy Storage*, vol. 136, p. 118596, 2025.

[5] K. L. Soon, N. S. Lai, C.-O. Chow et al., "A quantum-enhanced ensemble model to forecast battery state of health under varying discharging load," *Measurement*, vol. 256, p. 118318, 2025.

[6] J. Biamonte, P. Wittek, N. Pancotti et al., "Quantum machine learning," *Nature*, vol. 549, no. 7671, pp. 195-202, 2017.

[7] M. Schuld and N. Killoran, "Quantum machine learning in feature Hilbert spaces," *Phys. Rev. Lett.*, vol. 122, no. 4, p. 040504, 2019.

[8] V. Havlíček, A. D. Córcoles, K. Temme et al., "Supervised learning with quantum-enhanced feature spaces," *Nature*, vol. 567, no. 7747, pp. 209-212,

2019.

[9] M. AbuGhanem, "Superconducting quantum computers: who is leading the future?," *EPJ Quantum Technol.*, vol. 12, no. 1, p. 102, 2025.

[10] M. AbuGhanem and H. Eleuch, "NISQ computers: a path to quantum supremacy," *IEEE Access*, vol. 12, pp. 102941-102961, 2024.

[11] X.-D. Xie, X. Zhang, B. Koczor et al., "Advances in Quantum Computation in NISQ Era," *Entropy*, vol. 27, no. 10, p. 1074, 2025.

[12] J. Preskill, "Quantum computing in the NISQ era and beyond," *Quantum*, vol. 2, p. 79, 2018.

[13] M. Cerezo, A. Arrasmith, R. Babbush et al., "Variational quantum algorithms," *Nat. Rev. Phys.*, vol. 3, no. 9, pp. 625-644, 2021.

[14] M. Doyle, T. F. Fuller, and J. Newman, "Modeling of galvanostatic charge and discharge of the lithium/polymer/insertion cell," *J. Electrochem. Soc.*, vol. 140, no. 6, pp. 1526-1533, 1993.

[15] H. He, R. Xiong, and J. Fan, "Evaluation of lithium-ion battery equivalent circuit models for state of charge estimation by an experimental approach," *Energies*, vol. 4, no. 4, pp. 582-598, 2011.

[16] M. Berecibar, I. Gandiaga, I. Villarreal et al., "Critical review of state of health estimation methods of Li-ion batteries for real applications," *Renew. Sustain. Energy Rev.*, vol. 56, pp. 572-587, 2016.

[17] G. Nuroldayeva, Y. Serik, D. Adair et al., "State of health estimation methods for lithium-ion batteries," *Int. J. Energy Res.*, vol. 2023, no. 1, p. 4297545, 2023.

[18] Y. Li, K. Liu, A. M. Foley et al., "Data-driven health estimation and lifetime prediction of lithium-ion batteries: A review," *Renew. Sustain. Energy Rev.*, vol. 113, p. 109254, 2019.

[19] B. Bairwa, K. Pareek, and V. K. Jadoun, "Cycle based state of health estimation of lithium ion cells using deep learning architectures," *Sci. Rep.*, vol. 15, no. 1, p. 37078, 2025.

[20] S. Suh, D. A. Mittal, H. Bello et al., "Remaining useful life prediction of Lithium-ion batteries using spatio-temporal multimodal attention networks," *Heliyon*, vol. 10, no. 16, 2024.

[21] T. Alsuwian, S. Ansari, M. A. A. M. Zainuri et al., "A review of expert hybrid and co-estimation techniques for SOH and RUL estimation in battery management system with electric vehicle application," *Expert Syst. Appl.*, vol. 246, p. 123123, 2024.

[22] Y. Zhang, R. Xiong, H. He et al., "Long short-term memory recurrent neural network for remaining useful life prediction of lithium-ion batteries," *IEEE Trans. Veh. Technol.*, vol. 67, no. 7, pp. 5695-5705, 2018.

[23] K. Park, Y. Choi, W. J. Choi et al., "LSTM-based battery remaining useful life prediction with multi-channel charging profiles," *IEEE Access*, vol. 8, pp. 20786-20798, 2020.

[24] Y. Fan, Z. Lin, F. Wang et al., "A hybrid approach for lithium-ion battery remaining useful life prediction using signal decomposition and machine learning," *Sci. Rep.*, vol. 15, no. 1, p. 8161, 2025.

[25] L. Wang, S. Jiang, Y. Mao et al., "Lithium-ion battery state of health estimation method based on variational quantum algorithm optimized stacking strategy," *Energy Rep.*, vol. 11, pp. 2877-2891, 2024.

[26] C. Liang, S. Tao, X. Huang et al., "Stochastic state of health estimation for lithium-ion batteries with automated feature fusion using quantum convolutional neural network," *J. Energy Chem.*, vol. 106, pp. 205-219, 2025.

[27] S. Y.-C. Chen, S. Yoo, and Y.-L. L. Fang, "Quantum long short-term memory," *in Proc. IEEE Int. Conf. Acoust., Speech Signal Process. (ICASSP)*, 2022, pp. 8622-8626.

[28] A. Savitzky and M. J. E. Golay, "Smoothing and differentiation of data by simplified least squares procedures," *Anal. Chem.*, vol. 36, no. 8, pp. 1627-1639, 1964.

[29] M. Dubarry, C. Truchot, and B. Y. Liaw, "Synthesize battery degradation modes via a diagnostic and prognostic model," *J. Power Sources*, vol. 219, pp. 204-216, 2012.

[30] T. M. Cover, Elements of Information Theory. New York, NY, USA: *John Wiley & Sons,* 1999.

[31] H. Peng, F. Long, and C. Ding, "Feature selection based on mutual information criteria of max-dependency, max-relevance, and min-redundancy," *IEEE Trans. Pattern Anal. Mach. Intell.*, vol. 27, no. 8, pp. 1226-1238, 2005.

[32] G. Brown, A. Pocock, M.-J. Zhao et al., "Conditional likelihood maximisation: a unifying framework for information theoretic feature selection," *J. Mach. Learn.*

*Res.*, vol. 13, pp. 27-66, 2012.

[33] J. Zhu, Y. Wang, Y. Huang et al., "Data-driven capacity estimation of commercial lithium-ion batteries from voltage relaxation," *Nat. Commun.*, vol. 13, no. 1, p. 2261, 2022.

[34] K. A. Severson, P. M. Attia, N. Jin et al., "Data-driven prediction of battery cycle life before capacity degradation," *Nat. Energy*, vol. 4, no. 5, pp. 383-391, 2019.

[35] F. Xia, K. Wang, and J. Chen, "State of health and remaining useful life prediction of lithium-ion batteries based on a disturbance-free incremental capacity and differential voltage analysis method," *J. Energy Storage*, vol. 64, p. 107161, 2023.

[36] M. Schuld, A. Bocharov, K. M. Svore et al., "Circuit-centric quantum classifiers," *Phys. Rev. A*, vol. 101, no. 3, p. 032308, 2020.

[37] A. Kandala, A. Mezzacapo, K. Temme et al., "Hardware-efficient variational quantum eigensolver for small molecules and quantum magnets," *Nature*, vol. 549, no. 7671, pp. 242-246, 2017.

[38] K. Mitarai, M. Negoro, M. Kitagawa et al., "Quantum circuit learning," *Phys. Rev. A*, vol. 98, no. 3, p. 032309, 2018.

[39] J. R. McClean, J. Romero, R. Babbush et al., "The theory of variational hybrid quantum-classical algorithms," *New J. Phys.*, vol. 18, no. 2, p. 023023, 2016.

[40] F. Zou, L. Shen, Z. Jie et al., "A sufficient condition for convergences of Adam and RMSProp," *in Proc. IEEE/CVF Conf. Comput. Vis. Pattern Recognit. (CVPR)*, 2019, pp. 11127-11135.